\documentclass[prl,twocolumn,showpacs,amsmath,amssymb,showkeys]{revtex4}
\usepackage{graphicx}% Include figure files
\usepackage{bm}% bold math

\begin{document}

\title{Magnetotransport of lanthanum doped RuSr$_2$GdCu$_2$O$_8$ - the role of gadolinium}

\author{M.~Po\v{z}ek$^1$}
\email{mpozek@phy.hr} 
\author{A.~Dul\v{c}i\'{c}$^1$}
\author{A.~Hamzi\'{c}$^1$}
\author{M.~Basleti\'{c}$^1$}
\author{E.~Tafra$^1$}
\author{G.~V.~M.~Williams$^{2,3}$}
\author{S.~Kr\"{a}mer$^2$}
\thanks{\emph{Present address:}  Grenoble High Magnetic Field Laboratory, CNRS, B.P. 166, 38042 Grenoble, Cedex 9, France}                     
\affiliation{$^1$ Department of Physics, Faculty of Science, University of Zagreb, P. O. Box 331,
HR-10002 Zagreb, Croatia \\
$^2$ 2. Physikalisches Institut, Universit\"{a}t Stuttgart, D-70550 Stuttgart, Germany \\ 
$^3$ Industrial Research, P.O. Box 31310, Lower Hutt, New Zealand}

%\author{M.~Po\v{z}ek, A.~Dul\v{c}i\'{c}, A.~Hamzi\'{c}, M.~Basleti\'{c}, E.~Tafra}
%\affiliation{%
%Department of Physics, Faculty of Science, University of Zagreb, P. O. Box 331,
%HR-10002 Zagreb, Croatia
%}%

%\author{G.~V.~M.~Williams}
%\affiliation{
%2. Physikalisches Institut, Universit\"{a}t Stuttgart, D-70550 Stuttgart, Germany \\
%and Industrial 
%Research Limited, P.O. Box 31310, Lower Hutt, New Zealand.
%}%
%\author{S.~Kr\"{a}mer}
%\affiliation{
%2. Physikalisches Institut, Universit\"{a}t Stuttgart, D-70550 Stuttgart, Germany 
%}%

\begin{abstract}
Strongly underdoped RuSr$_{1.9}$La$_{0.1}$GdCu$_2$O$_8$ has been comprehensively studied by dc magnetization, microwave measurements, magnetoresistivity and Hall resistivity in fields up to 9 T and temperatures down to 1.75 K. Electron doping by La reduces the hole concentration in the CuO$_2$ planes and completely suppresses superconductivity. Microwave absorption, dc resistivity and ordinary Hall effect data indicate that the carrier concentration is reduced and a semiconductor-like temperature dependence is observed. Two magnetic ordering transitions are observed. The ruthenium sublattice orders antiferromagnetically at 155 K for low applied magnetic field and the gadolinium sublattice antiferromagnetically orders at 2.8 K. The magnetoresistivity exhibits a complicated temperature dependence due to the combination of the two magnetic orderings and spin fluctuations. It is shown that the ruthenium magnetism influences the conductivity in the RuO$_2$ layers while the gadolinium magnetism influences the conductivity in the CuO$_2$ layers. The magnetoresistivity is isotropic above 4 K, but it becomes anisotropic when gadolinium orders antiferromagnetically. 
\end{abstract}

\pacs{74.72.-h 74.25.Fy 74.25.Ha}
\maketitle

\section{Introduction}
\label{sec:level1}

The observation of magnetic order with a ferromagnetic (FM) component and superconductivity (SC) in the ruthenate cuprates is an intriguing issue that has motivated a number of studies especially since it was reported that superconductivity and the magnetic order coexist \cite{Bernhard:00}. It has been argued that the coexistence of the competing order parameters occurs via a spontaneous vortex phase \cite{Bernhard:00,Bernhard:99}, which is similar to the interpretation of the Ru1222 compounds \cite{Sonin:98}.

In RuSr$_2$RCu$_2$O$_8$ the low field magnetic ordering at $T_M \approx 133$ K is predominantly antiferromagnetic (AFM) with spin-canting leading to a small ferromagnetic component \cite{Lynn:00,Williams:00,Jorgensen:01,Takagiwa:01}. However, there is a spin reorientation with increasing magnetic field to a ferromagnetic phase \cite{Lynn:00,Williams:00,Takagiwa:01} that has been described as a spin-flop transition \cite{Williams:00}. It has recently been suggested that a small fraction of ferromagnetic nanoparticles appear dispersed in the antiferromagnetic lattice of RuSr$_2$RCu$_2$O$_8$ \cite{Cimberle:06}.

There is a general agreement that superconductivity is associated with the CuO$_2$ layers and magnetic order with the RuO$_2$ layers when R=Eu. It has been shown that both layers contain delocalized carriers \cite{Tokunaga:01} and coupling between the CuO$_2$ and RuO$_2$ layers is weak \cite{Pozek:02,McCrone:03}. Furthermore, nuclear magnetic resonance measurements \cite{Kramer:02} show that there is weak exchange coupling from Ru to Cu and electron paramagnetic resonance measurements show that there is also weak exchange coupling from Ru to Gd \cite{Fainstein:99}.

There is still an open debate concerning the nature of the low field antiferromagnetic order in the ruthenium sublattice.  Neutron scattering experiments suggest G-type antiferromagnetism where the Ru spins are aligned along the c-axis \cite{Lynn:00,Jorgensen:01} and zero-field nuclear magnetic resonance measurements suggest that the spins are aligned in the ab-plane \cite{Tokunaga:01}. In addition to the low-field AFM ordering of ruthenium sublattice, the gadollinium sublattice orders G-type antiferromagnetically at 2.8 K \cite{Lynn:00}. The dipolar fields from the AFM ordered ruthenium sublattice do not exactly cancel at the gadollinium site because of the spin canting \cite{Jorgensen:01}.

The coupling between the magnetic ordering in the Ru- and Gd-sublattices was indicated from magnetization measurements \cite{Williams:00,Jorgensen:01}, but the effect of this coupling on the transport properties was not studied because the pure compound is superconducting at the temperatures of interest.

In order to study the magnetic ordering at all temperatures, and its impact on
the transport properties, one needs to eliminate the superconducting state. 
For this reason, we have chosen to measure a La-doped sample, RuSr$_{1.9}$La$_{0.1}$GdCu$_2$O$_8$ (5\% of La on the strontium site). With this substitution, the lattice parameters only change by a very small amount, but the hole concentration in the CuO$_2$ planes is reduced when La$^{3+}$ replaces Sr$^{2+}$ and results in a hole concentration that is below that for superconductivity (less that 0.05 holes per Cu) \cite{Mandal:02}.

\section{Experimental Details}

The RuSr$_{1.9}$La$_{0.1}$GdCu$_2$O$_8$ ceramic sample was prepared as already described \cite{Williams:03}.
 
Resistivity, magnetoresistivity and Hall effect measurements were done using the standard six-contact configuration using the rotational sample holder and the conventional ac technique (22~Hz, 1~mA), in magnetic fields up to 9 T. 
The magnetoresistivity was measured with magnetic field (${\bf H}$) and current (${\bf I}$) in both, transversal (${\bf H}\  \bot \  {\bf I}$) and longitudinal (${\bf H} \parallel {\bf I}$) configurations.
Temperature sweeps for the resistivity 
measurements were performed with carbon-glass and platinum thermometers, while magnetic 
field dependent sweeps were done at constant temperatures where the temperature was 
controlled with a capacitance thermometer. 
 
The characterization of the samples by both, dc and ac magnetization measurements was done
using a SQUID magnetometer. 

Microwave measurements were carried out in an elliptical $_e$TE$_{111}$ copper cavity
operating at 9.3~GHz. The sample was mounted 
on a sapphire sample holder and positioned in the cavity center where the microwave electric 
field has a maximum.  The temperature of the sample could be varied from liquid helium to room 
temperature. Measurements were made with dc magnetic fields of up to 8 T. The details of the detection 
scheme are given elsewhere \cite{Nebendahl:01}.  
The measured quantities were $1/2Q$, the total losses of the cavity loaded by the sample, 
and $\Delta f/f$, the relative frequency shift from the beginning of the measurement. 
They are simply related to the surface impedance of the material $Z_s$, which depends on the complex conductivity $\widetilde{\sigma}$ and complex relative permeability 
$\widetilde{\mu}_r$. The total microwave impedance comprises 
both nonresonant resistance and resonant spin contributions.

\section{Results and Analyses}

\begin{figure}
\includegraphics[width=20pc]{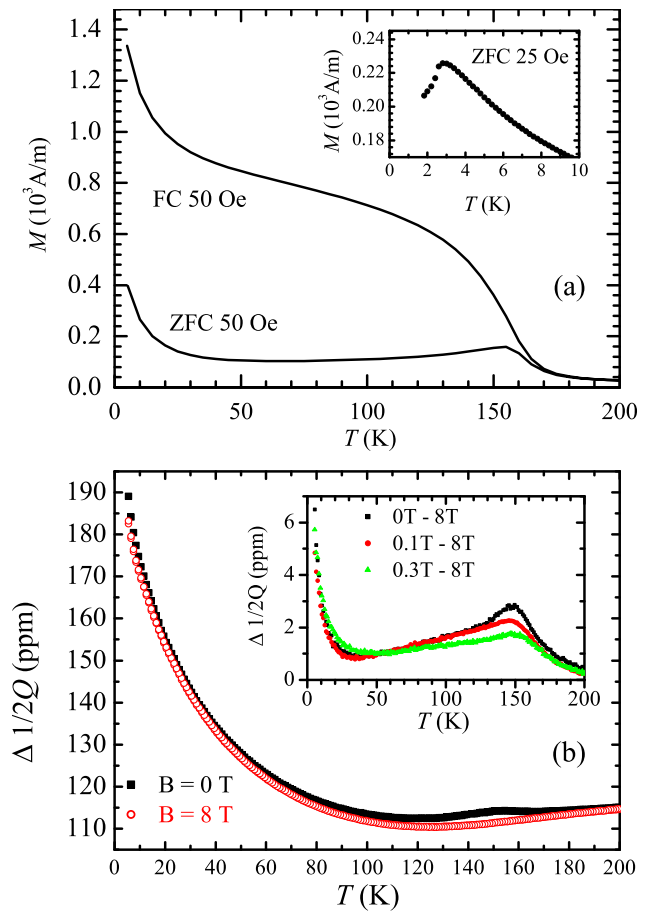}
\caption{(a) Field cooled (FC) and zero field cooled (ZFC) magnetization from the ceramic sample measured in a dc field of 50 Oe. Inset: ZFC magnetization at low temperatures, measured in dc field of 25 Oe; (b) Microwave absorption from the ceramic sample measured in zero field (full squares) and in $B=8$T (open circles). The inset shows the differences between the microwave absorption in various fields and the absorption at $B=8$T. The observed peak at 155 K disappears for fields higher than 1 T.} 
\label{fig_01}
\end{figure}

The dc magnetization of RuSr$_{1.9}$La$_{0.1}$GdCu$_2$O$_8$ is shown in Figure \ref{fig_01}a.
It is similar to the magnetization of the pure compound (shown in our 
previous work \cite{Pozek:02}) with a  
slightly higher magnetic ordering transition temperature of
$T_{Ru}\approx155$K arising from the RuO$_2$ planes, but without any sign of superconductivity in the CuO$_2$ planes down to 1.8 K.
The upturn in magnetization below 25 K is due to the onset of the magnetic order in the Gd sublattice with an antiferromagnetic transition at $T_{Gd}=2.8$K (inset). 

The temperature dependence of the microwave absorption in zero field and in $B=8$ T is shown in 
Figure  \ref{fig_01}b. The high temperature microwave absorption is about two times lar\-ger than in the
pure compound, indicating that the conductivity is lower in the La-doped compound. The zero field microwave absorption shows a peak at $T_{Ru}\approx155$ K which disappears at higher fields (inset). This behaviour is qualitatively and quantitatively equivalent to that observed in the pure compound \cite{Pozek:02}. At lower temperatures 
the absorption rises in contrast to the absorption in the superconducting parent compound.
This rise is a combined effect due to the sample resistivity and paramagnetic resonance from the Gd ions. 
\begin{figure}[t]
\includegraphics[width=18pc]{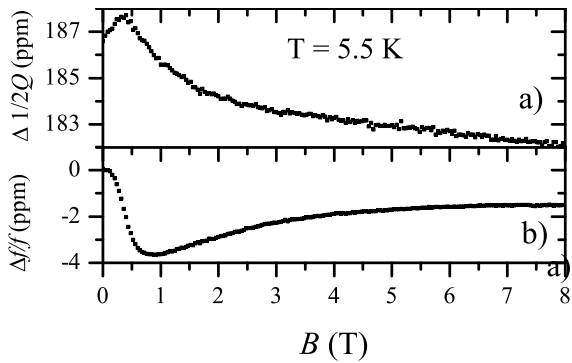}
\caption{Magnetic field dependence of the complex frequency shift at $T=5.5$ K:
(a) imaginary part of the complex frequency shift (absorption); (b) real part of the complex frequency shift (dispersion). The signals are dominated by electron spin resonance from the Gd$^{3+}$ ions.} 
\label{EPR}
\end{figure}%%
The magnetic field dependence of the microwave complex frequency shift at 5.5 K is shown in Figure \ref{EPR}.
The signal is dominated by electron paramagnetic resonance (EPR) from the Gd$^{3+}$ ions.
%, although the measurements were 
%done without the field modulation technique usually used in electron paramagnetic resonance (EPR) measurements. 
The EPR line is wide and the signal becomes detectable below 45 K.
This gadollinium EPR signal was obscured by the effects of superconducting weak links in the bulk pure compound, 
but it was observable in the powdered sample of the same pure compound \cite{Pozek:02}.

\begin{figure}
\includegraphics[width=18pc]{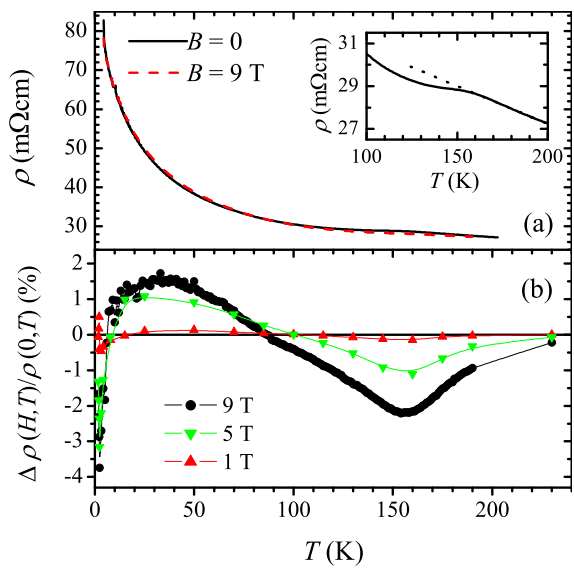}
\caption{(a) Resistivity in zero field and at $B=9$T; The inset shows the zero field resistance in an enlarged scale. The dotted line is the extrapolation of the temperature dependence of the resistivity from higher temperatures.(b) Transversal magnetoresistivity at three different fields. The lines are guides to the eye.} 
\label{RvsT}
\end{figure}

The resistivity in zero field and in $B=9$ T is shown in Figure  \ref{RvsT}a. 
At high temperatures the resistivity is roughly two times larger than in the pure compound. 
There is a kink in the zero field resistivity at 155 K that corresponds to the peak in the microwave absorption, and zero field cooled (ZFC) magnetization curves. It is the sign of the antiferromagnetic ordering in the RuO$_2$ planes.
The relative transversal magnetoresistivity $\frac{{\Delta \rho (H,T)}}{{\rho (0,T)}}$ at applied magnetic field of 1 T, 5 T, and 9 T, where $\Delta \rho (H,T) =\rho(H,T) - \rho (0,T)$,  is shown in Figure  \ref{RvsT}b. It can be seen that the magnetoresistivity at 9 T shows a pronounced minimum at 155 K, becomes positive at 85 K, shows a maximum at 30 K, and becomes negative again below 10 K. 
Note that the same behaviour is also present in the microwave absorption,
except for the sign change due to the inverted subtraction scheme in the inset to Figure \ref{fig_01}b. 
We aim to study in detail this complex temperature behaviour.
 
\begin{figure}[t]
\includegraphics[width=18pc]{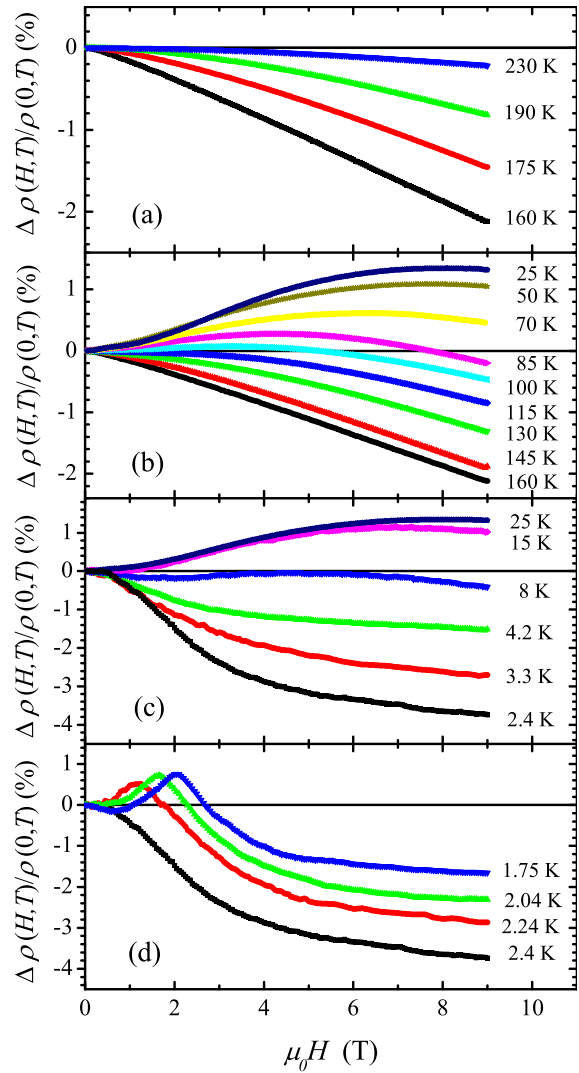}
\caption{Relative transversal magnetoresistivity:
(a) above 160 K; 
(b) between 25 K and 160 K;
(c) between 2.4 K and 25 K;
(d) below 2.4 K.} 
\label{MR}
\end{figure}

The magnetic field dependence of the relative transversal magnetoresistivity 
in a large temperature range from 230 K down to 1.75 K is shown in Figure \ref{MR}. 
The curves are grouped in subsets according to the physical processes that dominate in
each of the temperature ranges.

Figure \ref{MR}a shows the behaviour at high temperatures from 230 K down to 160 K. 
In this temperature range, there are ferromagnetic spin fluctuations without any long range magnetic order.
The application of an external field leads to a net thermal average moment and a reduction in the carrier scattering that results in a negative magnetoresistivity.

Antiferromagnetic long range order appears below 155 K in zero field, or very small applied fields, as can be seen in the magnetization curves in Figure \ref{fig_01}a. 
With the onset of antiferromagnetic order, the magnetoresistivity curves in Figure \ref{MR}b
become progressively less negative.
This behaviour is opposite to that in Figure \ref{MR}a, and indicates that a different physical process 
starts to play a role. 
%The magnetoresistivity curves monitor only the difference relative to the zero field.
%Hence, all the curves in Figure \ref{MR}b start from the zero level. 
For the interpretation of the MR 
curves below the magnetic ordering temperature, it is useful to look first at the behaviour of the zero field resistivity  (inset to Figure \ref{RvsT}a).
%and/or the microwave absorption in Figure \ref{fig_01}b. The zero field curves in those figures show that
%below the 
When the AFM order sets in, the resistivity drops below the values that might have been expected from the 
extrapolation of the temperature dependence observed above the transition (dotted line in the 
inset of Figure \ref{RvsT}a). 
Similar behaviour is also seen in the zero field microwave absorption in Figure \ref{fig_01}b. 
Since no appreciable change of the measured Hall coefficient is observed in the vicinity of 155 K (see later in this paper), we believe that the presently described observations can be interpreted as
clear evidence that the carrier scattering is reduced due to the onset of the AFM order. 
Having this in mind, we can now turn our attention again to the magnetoresistivity curves in Figure \ref{MR}b.
The increasing field first reduces the AFM order parameter, thus yielding an increase in the carrier scattering.
Therefore, the magnetoresistivity curve at 145 K shows a relative increase with respect to the 160 K curve.
When the long range AFM order is destroyed, the spin system is in the paramagnetic phase,
and the increasing applied field favours ferromagnetic fluctuations. This explains why the magnetoresistivity remains negative.
\begin{figure}
\includegraphics[width=18pc]{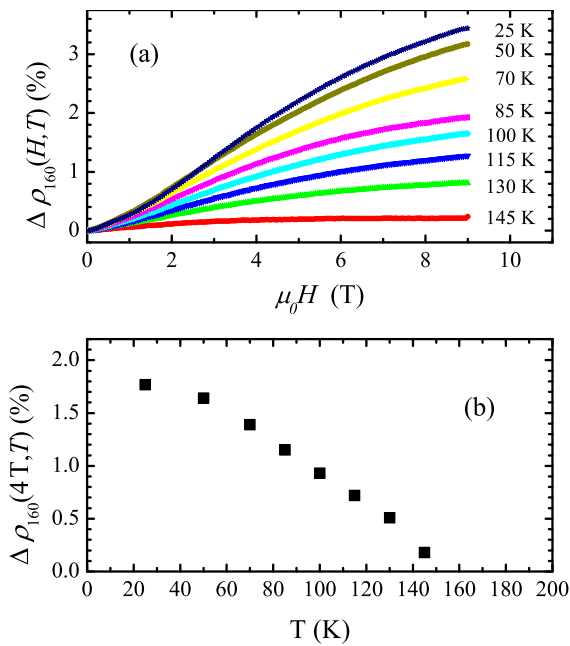}
\caption{(a) Net magnetoresistivity at temperatures between 25 K and 145 K (the curve at 160 K is subtracted). We denote $\Delta \rho _{160} (H,T) = \frac{{\Delta \rho (H,T)}}{{\rho (0,T)}} - {\frac{{\Delta \rho (H,160{\rm K})}}{{\rho (0,160{\rm K})}}} $. ; 
(b) The net magnetoresistivity at 4 T for temperatures below 160 K.} 
\label{phaseRu}
\end{figure}
However, as the temperature is lowered further, the AFM order parameter becomes larger. Below 100 K, it is strong enough that the initial 
rise of the magnetoresistivity curves dominates in Figure \ref{MR}b. 
%It takes large applied fields to reduce the AFM order parameter. 
Larger applied fields are required to reduce the AFM order parameter.

It is also known that an increased magnetic field can lead to spin canting,
which yields a FM component. Since the growing FM component tends to decrease the carrier scattering, the
magnetoresistivity curves would acquire a negative slope at high enough fields.
It is difficult to distinguish between this long range FM component and the field induced FM fluctuations 
that remain after any long range order parameter is destroyed. 
However, we observe in Figure \ref{MR}b that the negative slopes at the highest fields at lower temperatures are not bigger 
in absolute values than the slope of the curve at 160 K.
It seems that the long range FM component due to canting of the AFM order does not exceed the
field induced FM fluctuations.

We believe that it is possible to separate AFM and FM contributions to the magnetoresistivity, at least in an approximate way.
We have subtracted MR curve at 160 K 
from MR curves at lower temperatures, and the result is shown in Figure \ref{phaseRu}a. 
If one assumes that the 
FM contribution to the magnetoresistivity does not change appreciably below 160 K, the plotted curves represent an AFM contribution to the magnetoresistivity due to the ruthenium sublattice. The AFM order parameter in the ruthenium sublattice is gradually reduced with increasing 
magnetic field, and the introduced disorder makes a positive contribution to the magnetoresistivity.
At lower temperatures, the zero field AFM order parameter becomes larger, so that the corresponding curves in Figure \ref{phaseRu}a also show a larger rise. 
At high enough fields, when the AFM order is 
completely destroyed, the curves in Figure \ref{phaseRu}a saturate.
%This is further evidence which can be used for the resolution and interpretation of the
%even more complicated MR curves below 25 K.

The evolution of the ruthenium AFM order parameter can be followed in Figure \ref{phaseRu}b,
where the AFM contribution to the MR at 4 T is plotted.
This order parameter practically saturates at lower temperatures, and it is 
reasonable to assume that the ruthenium contribution to the MR curves does not change appreciably at 
temperatures below 25 K. The same conclusion would be reached if points for any other field values were plotted.
\begin{figure}[t]
\includegraphics[width=18pc]{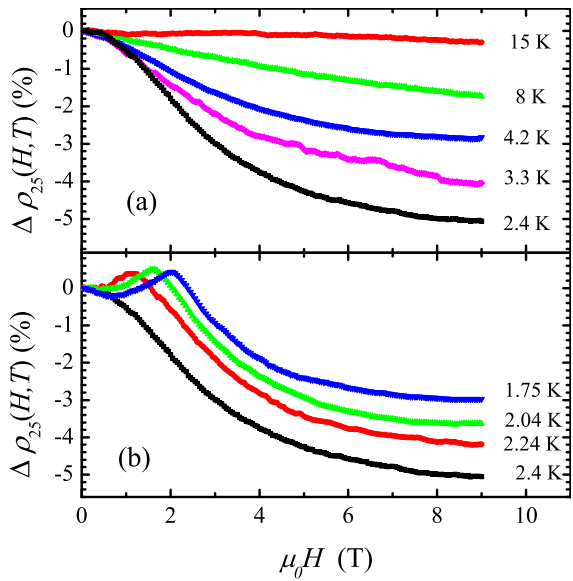}
\caption{Differences between transversal magnetoresistivity ($\Delta \rho _{25} (H,T) = \frac{{\Delta \rho (H,T)}}{{\rho (0,T)}} - {\frac{{\Delta \rho (H,25{\rm K})}}{{\rho (0,25{\rm K})}}} $) at various temperatures (The curve at 25 K is subtracted): 
(a) Net magnetoresistivity at temperatures between 2.4 K and 15 K;
(b) Net magnetoresistivity at temperatures between 1.75 K and 2.4 K. } 
\label{MRdiff}
\end{figure}

The analysis of the magnetoresistivity curves 
at still lower temperatures yields yet another interesting and new feature below 25 K (c.f. Figure \ref{MR}c).
%An interesting new feature appears below 25 K as seen in . 
The magnetoresistivity curves 
at 15 K (slightly) and 8 K (more pronounced) show that another negative contribution sets in and superimposes on the total signal.
It becomes dominant as the temperature is lowered down to 2.8 K.
Given the saturation trend observed in Figure \ref{phaseRu}, it is unlikely that a dramatic change occurs in the ordering of the ruthenium subsystem below 25 K. 
On the other hand, gadolinium spins exhibit enhanced paramagnetism below 25 K, as seen from the magnetization curves 
in Figure \ref{fig_01}a, and the microwave absorption in the inset to Figure \ref{fig_01}b. 
The paramagnetism from the Gd ions is so strong at 5.5 K that electron spin resonance can be observed (Figure \ref{EPR})
even without the common field modulation technique.
Hence, we ascribe the observed negative contribution 
to the magnetoresistivity in Figure \ref{MR}c to a
precursor of the AFM ordering of the Gd spin subsystem.
To distinguish the evolution of the gadolinium spin subsystem contribution from the already saturated ruthenium 
contribution, we can subtract the MR curve at 25 K from MR curves at lower temperatures. The results are shown in 
Figure  \ref{MRdiff}. 
The evolution above 2.8 K (Figure \ref{MRdiff}a) is qualitatively similar to the 
behavior of the Ru subsystem above 160 K observed in Figure \ref{MR}a. 
The gadolinium spin subsystem exhibits strong paramagnetism as a precursor to AFM ordering. The application of 
an external field stimulates parallel alignment of the Gd spins, thus reducing spin disorder and carrier scattering.
The largest negative magnetoresistivity in Figure \ref{MRdiff} is observed at 2.4 K, slightly below
the AFM ordering temperature of the Gd spin subsystem.

\begin{figure}
\includegraphics[width=18pc]{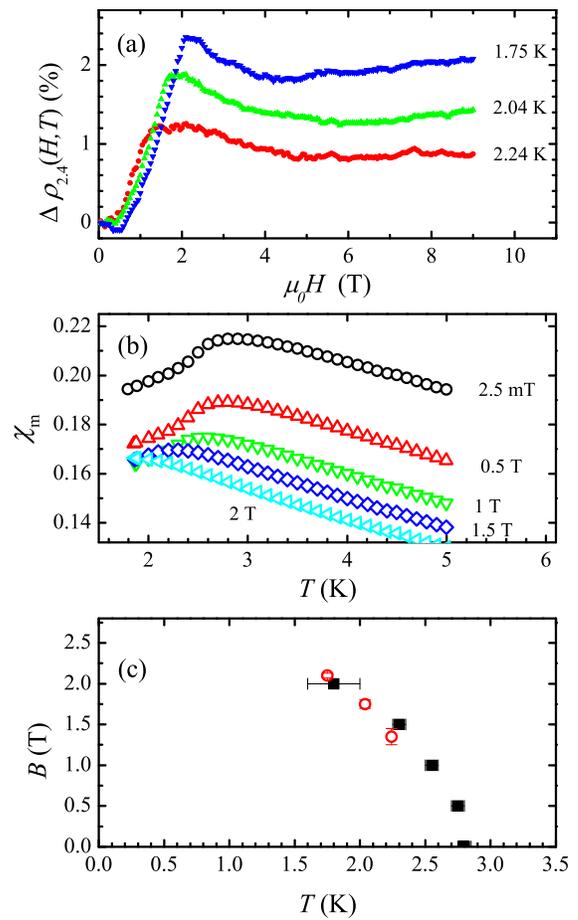}
\caption{ 
(a) Net magnetoresistivity at temperatures below 2.4 K 
when  magnetoresistivity at 2.4 K is subtracted ($\Delta \rho _{2.4} (H,T) = \frac{{\Delta \rho (H,T)}}{{\rho (0,T)}} - {\frac{{\Delta \rho (H,2.4{\rm K})}}{{\rho (0,2.4{\rm K})}}} $);
(b) Low-temperature magnetic susceptibility for several applied magnetic fields; 
(c) Peak positions of the magnetic susceptibility 
in Figure  \ref{Gd_phase}b (full squares) and saturation fields of the AFM magnetoresistivity in 
Figure  \ref{Gd_phase}a (open circles).} 
\label{Gd_phase}
\end{figure}

The effect of the increasing AFM order parameter at still lower temperatures is seen in Figure \ref{MRdiff}b.
Qualitatively, this behaviour is similar to that of the ruthenium subsystem below 155 K.
The initial rise of the magnetoresistivity is due to the 
destruction of the long range AFM order of the Gd subsystem. 
At higher fields, the negative component of the magnetoresistivity prevails.
In analogy with the ruthenium subsystem, we assume that the negative magnetoresistivity is due to the field induced FM 
fluctuations, and assume that this component does not change appreciably below 2.4 K.
Hence, we subtract the curve at 2.4 K 
from the MR curves at lower temperatures, and the result is shown in Figure \ref{Gd_phase}a. 
The initial rise of the magnetoresistivity appears to be analogous to the behaviour already seen
in Figure \ref{phaseRu}a for the ruthenium governed magnetoresistivity.

In order to check the correspondence of the magnetoresistivity to the phase diagram of the Gd spin sublattice, 
we have performed a series of magnetization measurements at low temperatures. 
The magnetic susceptibility at various applied fields is shown in
Figure  \ref{Gd_phase}b. The AFM transition temperature is suppressed by the magnetic field, 
and completely disappears in fields higher than 2 T. The peak positions are plotted in the phase 
diagram shown in Figure  \ref{Gd_phase}c. The field values where the AFM contribution to the magnetoresistivity reaches a maximum in 
Figure  \ref{Gd_phase}a, are also plotted in Figure \ref{Gd_phase}c as open symbols. 
It is obvious that the two observed features are well correlated.
Therefore we can identify the maxima in Figure \ref{Gd_phase}a with the field induced transition from the antiferromagnetic
to paramagnetic phase of the Gd spin subsystem.

\begin{figure}
\center
\includegraphics[width=18pc]{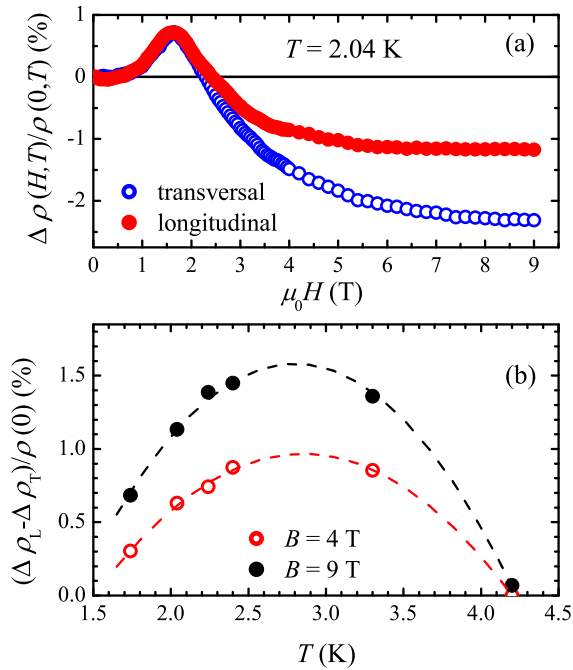}
\caption{(a) Transversal and longitudinal magnetoresistivity at $T=2.04$ K; (b) Temperature dependence of anisotropy at $B=4$ T (opened circles) and $B=9$ T (full circles). The plotted parabolas serve as guide to eyes.} 
\label{AMR}
\end{figure}

In order to extract additional information from the magnetoresistivity, we have measured it in both, longitudinal (${\bf H} \parallel {\bf I}$) and transversal 
(${\bf H}\  \bot \  {\bf I}$) configurations, so that the anisotropy can be determined.
The detected anisotropy in the magnetoresistivity is very small for temperatures above 4 K, similar to the observation in pure compound \cite{Pozek:02}. 
However, below 4 K a significant anisotropy appears between the transversal and longitudinal magnetoresistivity. 
The two curves taken at $T=2.04$ K are shown in Figure \ref{AMR}a. 
The anisotropy at other temperatures below 4 K is qualitatively similar to that in Figure \ref{AMR}a. 
Their common feature is that the
anisotropy does not show up at low magnetic fields where the magnetoresistivity is positive. This is the region where the AFM order of the Gd spins prevails. The anisotropy only gradually develops at higher fields. 
Since anisotropy in magnetoresistivity is a common feature of ferromagnetism, we gain yet another confirmation that ferromagnetic fluctuations develop at higher fields in the Gd spin subsystem.

The evolution of the anisotropy with temperature is also interesting. Figure \ref{AMR}b shows data taken at 4 T and 9 T.
The anisotropy reaches its maximum around $T_{Gd}=2.8$ K. Obviously, it corresponds to the gadolinium magnetic ordering temperature. It should be noted that anisotropy is usually not expected for the Gd$^{3+}$ ion whose half filled $f$-shell has a spherical charge distribution \cite{Campbell:82}.

\begin{figure}
\includegraphics[width=18pc]{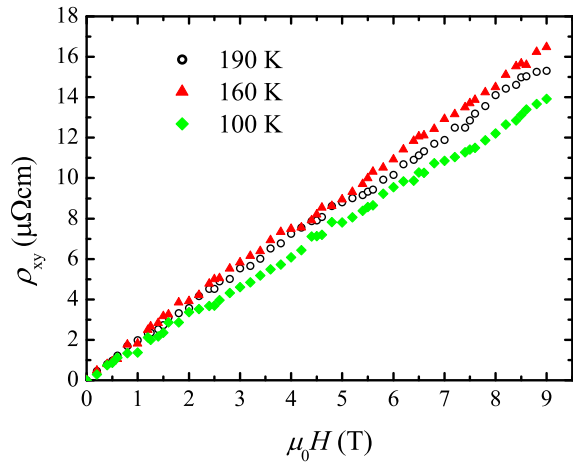}
\caption{Hall resistivity at three different temperatures.} 
\label{Hall}
\end{figure}

The Hall resistivity has also been measured and 
data for some high temperatures are shown in Figure \ref{Hall}. The Hall resistivity is linear up to 9 T for all measured temperatures, and the deduced Hall constant is very weakly temperature dependent.  Its value was roughly two times 
%the results are shown in . The high field Hall constant was roughly two times 
larger than in the undoped sample, indicating a reduced number of carriers. We note that in this sample there is no clear 
evidence of an extraordinary Hall effect. 
In our previous paper, the existence of extraordinary Hall effect was important
evidence that the RuO$_2$ layers are conducting in the pure sample \cite{Pozek:02}. The lack of the extraordinary 
Hall effect in the La-doped sample is most likely
due to the smaller and more linear magnetization in this sample, and does not prove the absence of conduction in the RuO$_2$ layers. 
On the contrary, we believe that the present MR data strongly indicate that both the RuO$_2$ and CuO$_2$ layers are conducting. 

%\begin{figure}
%\center
%\includegraphics[width=10pc]{RuGdColorSpins.eps}
%\caption{Crystal structure of RuSr$_2$GdCu$_2$O$_8$ .} 
%\label{unit_cell}
%\end{figure}

\section{Discussion}

The extensive measurements carried out in this work on La-doped Ru-1212:Gd sample have yielded complex results. The interpretation of the variety of features in the experimental data should be related to the crystal structure of RuSr$_2$GdCu$_2$O$_8$.
% (Figure  \ref{unit_cell}). One may observe that 
The ruthenium magnetic ions are relatively far away from the conducting CuO$_2$ planes. 

As pointed out by Picket et al.\cite{Picket:99}, the ruthenium magnetization lies within the $t_{2g}$ orbitals that do no directly couple to the Cu $d_{x^2-y^2}$ or Cu $s$ orbitals. Hence, their influence on the conduction in the CuO$_2$ layers is small, but they certainly influence the conduction in their own RuO$_2$ planes.  

Almost all of the magnetoresistivity above 25 K, seen in Figure \ref{MR}a and Figure \ref{MR}b is due to processes in the RuO$_2$ planes, which are both, conducting and magnetic. 

On the other hand, gadolinium magnetic ions are close to the conducting CuO$_2$ planes. Strongly localized gadolinium $f$ states do not significantly influence the 
relaxation rates of conduction electrons
in the distant RuO$_2$ layers but may have some effect
%their magnetic ordering, or disordering, can bring about a small scattering potential 
in the nearby conducting copper planes. Therefore, we ascribe the evolution of the magnetoresistivity below 25 K, seen in Figure \ref{MRdiff}, to scattering processes in the CuO$_2$ planes.

It is worth noting that the magnetoresistivity at all temperatures is a weak effect, and at most a few percent. This indicates that the ordering of the ruthenium sublattice at higher temperatures does not significantly affect the scattering rate of the carriers, probably because the scattering is already strong. 
%At lower temperatures, one would expect a large magnetoresistivity if the gadolinium ions were in the conducting planes. 
%Also, the magnetoresistivity would, in that case, lack anisotropy because of the spherical nature of Gd ion.
%However, the present experimental results are different, and can be interpreted with the gadolinium ions being out of the %conducting planes, and therefore causing just a small perturbation to the potential in the CuO$_2$ layers.
%As for the observed anisotropy in the magnetoresistivity, 
Gadolinium spin ordering at low temperatures has a small effect on magnetoresistivity because gadolinium is not embeded in the conducting plane.

Anisotropy in the magnetoresistivity is not expected for the Gd$^{3+}$ ion with spherical charge distribution \cite{Campbell:82}.
We may tentatively ascribe the presently observed anisotropy to the side position of the gadolinium ion with respect to the CuO$_2$ plane, so that some asymmetry is induced.
%not seen from the perspective of the conducting plane.

\section{Conclusions}

In conclusion, the replacement of 5\% Sr$^{2+}$ ions by La$^{3+}$ ions in RuSr$_2$GdCu$_2$O$_8$ further decreases doping in the CuO$_2$ planes in the already underdoped parent compound to a level where superconductivity is completely suppressed. The number of carriers is strongly reduced as revealed by the Hall resistance.Dc and microwave resistivities are two times larger than in the pure compound at higher temperatures and show semiconductor-like behaviour at lower temperatures. 

However, there still remain two conducting layers that are effectively decoupled. The conducting RuO$_2$ layers are influenced by magnetic ordering of the Ru spins as already observed in the parent compound. We detect and explain the influence of gadolinium magnetism on the conductivity. 
Gadolinium localized spins do not alter the electronic band structure of the CuO$_2$ layers, but may influence the relaxation rates of normal-state electrons when superconductivity is destroyed by other means (underdoping).

\section*{Acknowledgments}

We acknowledge funding support from the Croatian Ministry of Science and Technology, the 
New Zealand Marsden Fund, the New Zealand Foundation for Research Science and Technology, and the Alexander von Humboldt Foundation.
MP thanks to Dr. Ivan Kup\v{c}i\'c for valuable discussions.

\end{document}